\documentstyle[prl,aps,multicol,epsfig,epsf]{revtex}
\begin{document}
\draft
\title{Klein Factors in multiple Fractional Quantum Hall edge
tunneling}
\author{R. Guyon$^1$, P. Devillard$^{1,2}$, T.
Martin$^{1,3}$ and I. Safi$^{1}$}
\address{$^1$ Centre de Physique Th\'eorique, Case 907 Luminy,
13288 Marseille Cedex 9, France}
\address{$^2$ Universit\'e de
Provence, 13331 Marseille Cedex 03, France}
\address{$^3$ Universit\'e de la M\'editerran\'ee,
13288 Marseille Cedex 9, France} 
\maketitle
\begin{abstract}
A fractional quantum Hall liquid with multiple edges is
considered. The computation of transport quantities 
such as current, noise and noise cross correlations
in such multiple edge samples 
requires the implementation of so called Klein factors,
which insure the correct quasiparticle exchange
properties. The commutation relations of these factors
are obtained by closing the system into a single edge. 
The non-equilibrium  Green's function 
formalism associated with such factors is derived for a 
simple Laughlin fraction of the Hall effect.
It is shown explicitly how Klein factors enter the calculation
of the noise cross correlations, as well as the 
correction to the Poisson limit for the noise.  
\end{abstract} 
 
\begin{multicols}{2}
\narrowtext

The edge state picture of the fractional quantum Hall 
effect \cite{laughlin_stormer} (FQHE) has led to fundamental
discoveries. Various systems have been studied
to analyze the 
quasiparticle excitations which
tunnel through 
 such quantum fluids \cite{kane_fisher,chamon_noise}. 
In particular, 
for a point contact
 geometry
with two edges, the detection of the backscattering
noise leads to the measurement of the quasiparticle charge 
\cite{saminadayar}. 
At the same time, resonant antidot
tunneling geometries have allowed a capacitive
measurement of the charge \cite{goldman} in multiple edge
geometries. 
Aside from this situation, multi edge geometries
have not received much attention. Yet  
it was argued recently
 \cite{safi_devillard_martin}
that the detection of noise
correlations in a three edge system
constitutes a direct link to the 
quasiparticles statistics. One edge can
be depleted due to the tunneling of quasiparticles 
into other edges. The Hanbury-Brown 
and Twiss type geometry \cite{HBT} of Ref. 
\cite{safi_devillard_martin} probes whether two 
quasi-holes on the injecting edge 
are allowed to overlap. Here,
the key point is that
quasiparticle edge tunneling operators need 
to be supplemented by Klein factors in order to
specify the statistics. Klein factors were so far
mostly introduced to tackle fermionic problems 
\cite{nayak_gogolin}, 
yet it seems that their implementation is 
quite relevant for complex fractional 
Hall geometries. 

The purpose here is to give a pedagogical view of how 
fractional statistics enters these multiple edge geometries: 
which (fractional) phase is generated
when two quasiparticles  from different edges 
are exchanged ? We focus on Laughlin filling factors $\nu = (2p+1)^{-1}$ ($p$ integer). 
First, specifying a quasiparticle tunneling Hamiltonian, we
derive the algebra
 which is obeyed by Klein factors. Next, we
introduce the 
 Keldysh Green's functions for the Klein
factors. These tools are then applied to study the
non-equilibrium current and noise correlations in a
multi-edge system. In particular, it will 
be shown that Klein
factors are unnecessary for simple, two 
edge samples. Finally,
we argue that when considering the non-equilibrium average of a
tunneling operator between two edges, 
 there remains always a
contribution associated with Klein
 factors, which is
symptomatic of quasiparticle exchange with
 other edge
``reservoirs''.
 
\begin{figure}
 \epsfxsize 8 cm 
\centerline{\epsffile{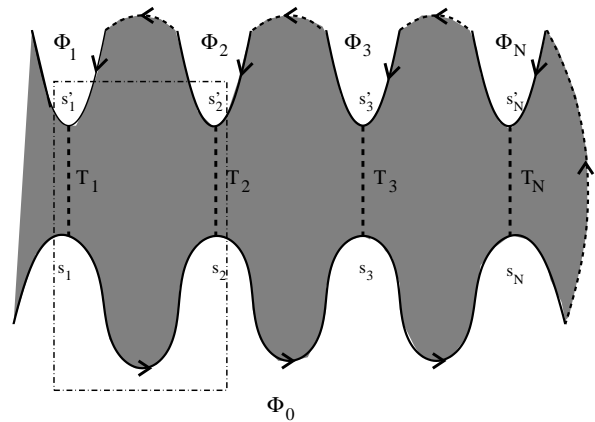}}
 \medskip 
\caption{\label{conect} Multiple edge geometry: tunneling
processes $T_i$ between the injecting edge specified by $\phi_0$
and other edges, with fields labelled $\phi_1$, $\phi_2$,...,
$\phi_N$. Curvilinear abscissas ($s,s'_1,s'_2,...$) are attached
to each edge. In our disconnected edge geometry only the full
lines defining the edges are kept. 
When considering the connected, single edge problem
the dotted lines are added to fix the topological 
constraints. Inside frame shows how the transport properties
in $T_2$ are affected by the fluctuating current in $T_1$.  
}   
\end{figure}
Our starting point is the Hamiltonian \cite{wen_edges} which
describes a system
 of decoupled edges
$H_{0}=(v_F\hbar/4\pi)\sum_{l=0}^{N}\int ds
(\partial_{s}\phi_l)^2$ with $s$ the curvilinear abscissa.
 The
fields 
 $\partial_{s}\phi_l$ enter the description of long
wave length excitations along each edge.
 
Tunneling between
edge $0$ and the other edges
 is described by the hopping Hamiltonian
\cite{chamon_noise}:
 \begin{equation}
 H_{t} = \sum_{l=1}^N \sum_{\epsilon=\pm} \Gamma_l   e^{-i \epsilon
e^*\chi_l/\hbar c} T_l^{(\epsilon)} \end{equation}
where $T_l = T_l^{(+)}=\psi_{0}^\dag(s_l)  \psi_{l}(s'_l)  =T_l^{(-)\dagger}$.
$\psi_{l}^\dag$ creates a quasiparticle excitation with 
charge $e^* = \nu e$  on edge $l$.
Here ``0'' refers to a specific edge where, for instance,
current can be injected into the others. 
A voltage bias can be imposed as usual using the Peierls
substitution, i.e. $e^* \chi_l(t) = \hbar \omega_0 c t$ for a constant DC bias
$V_l= \hbar \omega_0 /e^* $ imposed between edges $0$ and $l$.  
The quasiparticle operators are written in a bosonized
form  $\psi_l(s)= (2\pi\alpha)^{-1/2} e^{-i\sqrt{\nu}\phi_l(s)~}$
with $\alpha$ a short distance cutoff.
Note that no fractional statistics is taken into account when
quasiparticles from different edges are exchanged, whereas
\cite{fisher_glazman}:
\begin{equation}
\psi_l(s) \psi_l(s') = e^{\pm i \sqrt{\nu} \pi sgn(s-s')} \psi_l(s')
\psi_l(s) 
 \end{equation}
on a given edge. 
Fractional statistics is a consequence of the Laughlin
correlated state describing the FQHE.  
The commutation relations for the $\psi_l$'s should originate
from the derivation Chern--Simons boundary action. Here, we
conjecture that the commutation relations are similar to 
those for a system of connected edges.
For this reason, 
Klein factors are introduced, which means that each
quasiparticle operator $\psi_l$ is changed to $F_l\psi_l$. The
requirement of unitarity, $F_l^\dagger F_l=F_l F_l^\dagger=1$, 
is  consistent with the fermionic case 
($\nu=1$).

Consider a closed system
(Fig.\ref{conect})  where edges are connected with an open  contour
which brings all the edges into one. Klein factors are not
necessary in this system because the commutation relations 
of tunneling operators are enforced by the single
chiral bosonic field $\phi$.
Yet, by looking at the commutation
 relations of different
tunneling operators we show below 
 that for a system with the
same topology but where 
 edges are disconnected, the Klein 
factors can be computed.  
 The operator describing tunneling
from edge ``$0$'' (location
 $s_l$) to edge ``$l$'' (location
$s'_l$) reads $T_l = (2 \pi \alpha)^{-1} e^{i\sqrt{\nu}\phi(s_l)~}  
e^{-i\sqrt{\nu}\phi(s'_l)~}  $. 
 
Using the Baker--Campbell--Hausdorff formula \cite{baker}, 
one can compute the commutator of two tunnel 
operators:
\begin{equation}
[T_j, T_k]=
e^{-i\pi\nu}T_kT_j
\left[1-e^{-i\pi\nu[sgn(s_j-s_k)+sgn(s'_j-s'_k)]}\right] 
\end{equation}
Chirality imposes that 
$sgn(s_k-s_j)=-sgn(s'_j-s'_k)$ if the two tunneling paths
do not cross. In this convention, which will be kept  
throughout the paper, $[T_j,T_k]=0$. In the opposite
situation where the tunneling paths have a crossing, 
$T_jT_k= e^{2i\pi\nu sgn(s_j-s_k)}T_kT_j$.

We now turn to the same system (same topology) 
depicted in Fig. \ref{conect} but  where all edges are
disconnected.
In order to enforce fractional statistics, we
introduce Klein factors in such a way that the tunneling
operators become $T_l = \psi^\dag_0 (s_l) \psi_l(s'_l) F_0
F_l^\dag$.
 We require that the commutator of the tunneling
operators of this disconnected system give
the same results as the connected one.
The Klein factors thus have to satisfy the relation :

\begin{equation}
 \label{relKF}
 ( F_0 F_k^\dag) 
(F_0 F_j^\dag)
 =  e^{i \nu \pi sgn(s_k -s_j)} (F_0 F_j^\dag) 
(F_0 F_k^\dag)
 \end{equation}

Eq. (\ref{relKF}) is compatible with fractional
statistics, i.e. 
$F_j F_k = e^{- i \pi \nu p_{j k}} F_k F_j$, for $j\neq k$
where $p_{jk} =\pm 1$.  For arbitrary $j$ and $k$, $p_{jk}$  are
the elements of an antisymmetric $N\times N$ matrix with elements $0,1,-1$
which are tied to the edge configuration.  
Using Eq. (\ref{relKF}), we find a connection between the
elements $p_{jk}$ and the sign of the algebraic distance : 
$sgn(s_k -s_j) =  p_{k 0} + p_{0 j} + p_{jk}$.
 If for
instance $k=2$ and $j=1$ with $sgn(s_2 - s_1) =
 1$,
we can choose : $p_{01} = -1 = - p_{02} = - p_{12}$.  
 
These statistical constraints are now applied to a define
problem of fractional Hall edge transport: the computation of
current and noise correlations associated with several 
tunneling processes.  We return to  $N+1$ edges 
in the previous 
 system with $N$ tunnelings paths. We represent
the current on 
 lead $l$ using the Keldysh formulation of non-equilibrium
 transport:   
 \begin{equation}
 \langle I_l \rangle =
\frac{1}{2} \sum_{\eta = \pm} \langle T_K  I_l(t^\eta)
e^{-\frac{i}{\hbar} \int_K H_{t} (t') dt'}   \rangle
\label{current_keldysh}
\end{equation}
where $\eta$ labels the branch of the time ordering contour K. 
 
The cross current fluctuations
(i.e. the noise correlations) between 
two leads $j$ and $k$ are given by:
 
\begin{equation}
 S_{jk} (t,t') =  \sum_{\eta = \pm} \langle
T_K I_j(t^\eta)I_k(t'^{-\eta})   e^{-\frac{i}{\hbar}
\int_K H_{t} (t_1) dt_1}  \rangle
\label{noise_keldysh}
 \end{equation}
The perturbative computation of these operators is achieved by 
expanding the evolution operator. The finite frequency 
noise $S_{jk}(\omega) = \sum_n
S^{(n)}_{jk}(\omega)$ ($n^{th}$ order in $\Gamma_j$'s)
is the Fourier transform of the real time correlator (Eq.
(\ref{noise_keldysh})). The $(n+2)^{th}$ order of the noise
is :
  \begin{eqnarray}
  S_{j
k}^{(n+2)} (\omega) &=&\int\! dt \, e^{i \omega t} \Big\{ \frac{\Gamma_j^{(\epsilon_j)} \Gamma_k^{(\epsilon_k)}}{n!} \Big(-\frac{i}{\hbar}\Big) ^{n} 
\!\int dt_1\! ..dt_{n}  
\nonumber
\\
 &&\times \sum_{\eta \eta_1 .. \eta_{n}}   \sum_{\epsilon \epsilon'} 
\sum_{\epsilon_1...\epsilon_{n}} \epsilon 
\epsilon'
e^{i
\epsilon \omega_0 t - i \omega_0 \sum_{l = 1}^{n}
\epsilon_l t_l} 
 \nonumber \\
 &&\times \langle T_K
T_j^{(\epsilon)}(t^\eta) T_k^{(\epsilon')}(0^{-\eta}) 
\sum_{q_1 = 1}^{N} \Gamma^{(\epsilon_1)}_{q_1} T_{q_1} ^{(\epsilon_1)}
(t_1^{\eta_1}) .... \nonumber \\
  &&~~...\sum_{q_{n}
= 1}^{N} \Gamma^{(\epsilon_n)}_{q_{n}} T_{q_{n}} ^{(\epsilon_{n})}
(t_{n}^{\eta_{n}})  \rangle  - \langle I_j(t) \rangle 
 \langle I_k(0) \rangle \Big\}
 \end{eqnarray}
The
 lowest non-vanishing contribution to the noise in a given
terminal ($j=k$) is of order  $2$ in the tunneling amplitude, i.e.
the Poisson result. The cross correlations require to 
compute the $4^{th}$ order contribution. For example, the non--vanishing 
contribution to noise correlator between two terminals ($j<k$) :
 
\begin{eqnarray}
  S^{(4)}_{j k} (t,t') &=&
\frac{-2 e^{*2}}{\hbar^2} \Big(\frac{-i}{\hbar}\Big)^2
\sum_{\eta \eta_1
 \eta_2} \sum_{\epsilon \epsilon'
\epsilon_1 \epsilon_2} \epsilon
 \epsilon' \vert \Gamma_j
\Gamma_k \vert ^2 \eta_1 \eta_2 \int \!dt_1\! \int\!
 dt_2 \nonumber \\
&& ~ \langle
T_K  T_j^{(\epsilon)}
 (t^\eta)  T_k^{(\epsilon')}
(t^{' -\eta}) 
T_j^{(\epsilon_1)} (t_1^{\eta_1}) T_k^{(\epsilon_2)}
(t_2^{\eta_2})  \rangle \nonumber \\
&& \exp\left[i \epsilon \chi_j(t) + i \epsilon' \chi_k(t')
+i \sum_{l=j,k} \epsilon_l \chi_l (t_l) \right] 
\label{noise_time_fourth}
\end{eqnarray}
To this order, the Keldysh ordered product contains the product
of two contributions. One of such originates from the
computation of the average of the bosonic fields associated
with each edge. This contains the dynamical aspect of these
fields, and gives rise to exponentiated  chiral Green
functions. The other factor has no dynamics other than that
specified by time ordering; It contains the Klein factors: 
\begin{eqnarray}
&&  \langle
T_K \Big\{ T_j^{(\epsilon)} 
(t^\eta)  T_k^{(\epsilon')} (t^{' -\eta}) 
T_j^{(\epsilon_1)} (t_1^{\eta_1})
T_k^{(\epsilon_2)}  (t_2^{\eta_2}) \Big\} \rangle
= 
\nonumber
\\
&& 
(2 \pi \alpha)^{-4} \langle T_K 
e^{-i\sqrt{\nu} \epsilon \phi_j^{\eta}(0,t)} 
e^{-i\sqrt{\nu} \epsilon_1 \phi_j^{\eta_1}(0,t_1)}
\rangle\nonumber\\
&&
\times
\langle T_K 
e^{-i\sqrt{\nu}\epsilon' \phi_k^{-\eta}(0,t')} 
e^{-i\sqrt{\nu}\epsilon_2 \phi_k^{\eta_2}(0,t_2)}
\rangle\nonumber\\
&&
\times
\langle T_K 
e^{+i\sqrt{\nu}\epsilon\phi_0^{\eta}(-s_{kj}/2,t)}
e^{+i\sqrt{\nu}\epsilon'\phi_0^{-\eta}(s_{kj}/2,t')}
\nonumber\\
&&~~~ 
\times
e^{+i\sqrt{\nu}\epsilon_1\phi_0^{\eta_1}(-s_{kj}/2,t_1)}
e^{+i\sqrt{\nu}\epsilon_2\phi_0^{\eta_2}(s_{kj}/2,t_2)}
\rangle
\nonumber\\
&&
\times
\langle T_K 
(F_0 F_j^\dag)^{(\epsilon)}(t^{\eta})
(F_0 F_k^\dag)^{(\epsilon')}(t^{' -\eta})
\nonumber\\
&~&~~~~ 
\times
(F_0 F_j^\dag)^{(\epsilon_1)}(t_1^{\eta_1})
(F_0 F_k^\dag)^{(\epsilon_2)}(t_2^{\eta_2})
\rangle
~,
\label{ordered_tunnel}
\end{eqnarray}
where $s_{kj}=s_k-s_j$. The contraction of the chiral fields gives
rise to quasiparticle conservation laws
$\epsilon=-\epsilon_j$ and $\epsilon'=-\epsilon_k$ . It is most convenient 
to compute the time ordered Klein factor product 
using a bosonization formulation similar to 
that used for the quasiparticle fields:
\begin{equation}
F_0 F_k^\dag \equiv e^{-i \sqrt{\nu} \theta_k}~.
\label{bosonic_klein}
\end{equation}
Introducing the
 fact that  $ sgn(s_k -s_j) = 1$ in Eq.
\ref{relKF} , we find (using the
Hausdorff formula) that
 $[\theta_j,\theta_k] = i \pi$.
From the fields $\theta_i$, annihilation and creation 
operators $a$ and $a^\dagger$
can then be defined as $a
\equiv 
\frac{1}{\sqrt{2 \pi}} (\theta_j +i\theta_k)$,
with the usual property that 
$\langle a^\dag a \rangle_0 = 0 $ (``ground'' state)
and $\langle a a^\dag \rangle = 1$.

The ground state expectation value 
of products of the bosonic fields reads: $\langle \theta_j \theta_k \rangle_0 
= - \langle \theta_k \theta_j \rangle_0 = i \frac{\pi}{2}$ and 
$ \langle \theta_j \theta_j \rangle = \langle \theta_k \theta_k
\rangle = \frac{\pi}{2}$. 
 
Because we are dealing with a non-equilibrium transport
situation, a Keldysh matrix is introduced for 
the bosonic fields
of Eq.(\ref{bosonic_klein}) :
\begin{eqnarray}
g_{jk}^{\eta_1\eta_2}(t)&\equiv& \langle T_K
\Big\{
 \theta_j(t^{\eta_1}) \theta_k(0^{\eta_2})  \Big\}
\rangle\nonumber\\
&=&i
\frac{\pi}{2} \Big\{ \frac{\eta_1 + \eta_2}{2}\, sgn(t)
- \frac{\eta_1 - \eta_2}{2} \Big\} ~.
\end{eqnarray}
For the same tunneling events one finds 
$g_{l l}^{\eta_1 \eta_2} (t)  = \frac{\pi}{2}$. 

The application of these Klein factor Green's function
is most relevant when considering time ordered products
\begin{eqnarray}
&&\langle T_K  \Pi_{l = j,k}\; e^{i \epsilon_{l 1} 
\sqrt{\nu} \Theta_l (t_{l 1}^{\eta_{l 1}})} .....e^{i
\epsilon_{l m_l} \sqrt{\nu} \Theta_l (t_{l m_l}^{\eta_{l m_l}})}
 \rangle \nonumber\\
&&=\exp\bigl(- \nu \sum_{l \leq l'}  
\sum_{k=1}^{m_l}\sum_{k'=1}^{m_{l'}} \epsilon_{l
k}
 \epsilon_{l' k'} g_{l
l'}^{\eta_{l k} \eta_{l' k'}} (t_{l k} - t_{l' k'}) \bigr)  
\label{arbitrary}
\end{eqnarray}
($l,l'=j,k$)
which vanishes unless $\sum_{k=1}^{m_l} \epsilon_{l k} = 0$.
 
Consider first the case of a two edge system, where 
only one Klein factor is necessary. 
In Eq. (\ref{arbitrary}), the tunnel locations are $l=l'=1$, 
so that the only Green's function present        
is  $g_{1 1}^{\eta_1 \eta_2}(t)=\pi/2$.
Together with the requirement of Eq. (\ref{arbitrary}) to give 
a non-zero result, the term in the exponential then vanishes. 
This completes the proof that for a two edge system,
when computing non-equilibrium averages for the current, the
density fluctuations or the noise, the Klein factors do not 
contribute\cite{ketteman}, and this, order by order  in the perturbation
series.

We now turn to the 
computation of the zero frequency
current cross correlations between tunneling paths $1$ and $2$
in presence 
of a DC bias.
 Eq. (\ref{noise_time_fourth}) is
integrated
 over $t$: 
\begin{eqnarray}
&&\tilde{S}_{j k} (\omega = 0) = \frac{2 e^{*2}}{\hbar^4}
\sum_{\eta \eta_1  \eta_2} \sum_{\epsilon \epsilon'} 
\epsilon \epsilon' 
 \eta_1 \eta_2 \int \! dt \! \int \! dt_1 \! \int \! dt_2
\nonumber \\  && \frac{\vert \Gamma_j \Gamma_k \vert^2}{(2\pi\alpha)^4} \;
e^{i\omega_0 [ \epsilon (t - t_1) -\epsilon' t_2]} 
e^{2\nu\left[G_{\eta\eta_1}(0,t-t_1)+G_{-\eta\eta_2}(0,-t_2)\right]}
\nonumber
\\
&& \times
\Big( \frac{e^{\nu\epsilon \epsilon' \left[G_{\eta\eta_2}(-s_{kj},t-t_2)
+G_{-\eta\eta_1}(s_{kj},-t_1)\right]}}{e^{\nu\epsilon \epsilon'
\left[G_{\eta-\eta}(-s_{kj},t)
+G_{\eta_1\eta_2}(-s_{kj},t_1-t_2)
\right]}} e^{i \xi} -1 \Big)
\end{eqnarray}
Where the Klein factor time ordered product reads :
\begin{equation}
e^{i \xi} = 
\langle T_K 
e^{-i\epsilon \sqrt{\nu}\theta_j(t^{\eta})}
e^{-i\epsilon' \sqrt{\nu}\theta_k(0^{-\eta})}
e^{i\epsilon \sqrt{\nu}\theta_j(t_1^{\eta_1})}
e^{i\epsilon' \sqrt{\nu}\theta_k(t_2^{\eta_2})} \rangle.
\end{equation}

Note that the product of current averages $\langle I_j \rangle 
\langle I_k \rangle$ has to be subtracted before computing the Fourier 
transform. The resulting expression then identifies with the noise correlator
which is typically measured in experiments \cite{henny}. 
Because of the symmetry properties of the 
Keldysh Greens function, only the 
configuration $\eta_1=-\eta=-\eta_2$ 
needs to be considered and  the configuration
$\epsilon=-\epsilon'$ is chosen.
 Indeed, this turns  
out to be the relevant configuration which  
generates a series solution of the noise correlations 
\cite{safi_devillard_martin}.  Choosing the Klein factor
correlator with $\epsilon=+$:
 \begin{eqnarray}
&&\langle T_K 
e^{-i\sqrt{\nu}\theta_j(t^{\eta})}
e^{i\sqrt{\nu}\theta_k(0^{-\eta})}
e^{i\sqrt{\nu}\theta_j(t_1^{\eta_1})}
e^{-i\sqrt{\nu}\theta_k(t_2^{\eta_2})} \rangle 
= \nonumber\\
 &&
= \exp\left[ \nu \left( 
g_{jk}^{\eta -\eta}(t)+
g_{jj}^{\eta -\eta}(t-t_1)-
g_{jk}^{\eta \eta}(t-t_2)
\right)\right] 
\nonumber\\
&&
\times\exp\left[ \nu \left( 
-g_{kj}^{-\eta -\eta}(-t_1)+
g_{kk}^{-\eta \eta}(-t_2)+
g_{jk}^{-\eta \eta}(t_1-t_2)
\right) \right]
\nonumber
\\
&&
=\exp\left[ \frac{i \nu \pi}{2}\,\eta 
\left(sgn(-t_1)-sgn(t-t_2)
\right) \right]
\label{config_+}
\end{eqnarray}
The Klein factor contribution with 
$\epsilon=-$ is simply the complex conjugate of Eq. 
(\ref{config_+}). This term is crucial for a correct
description of the noise cross correlations in the FQHE.
For $\nu=1$, it allows to recover the negative correlation
result obtained with scattering theory
\cite{landauer_martin}. For $\nu=1/3$, the noise correlations
exhibit an anti-bunching behavior, which is weaker than
the integer case. 

Finally, we turn to the computation of
non-equilibrium  averages attached to a given lead: 
the zero frequency current fluctuations
in lead $1$, denoted $\tilde{S}_{j}(\omega=0)$.
Perturbation theory tells us that there are two 
contributions to the current and noise: 
\begin{eqnarray}
\langle I_{j} \rangle &=& \langle I_{j} \rangle_0+
\langle I_{j}\rangle_K
\\
\tilde{S}_{j}&=&\tilde{S}_{j}^0+\tilde{S}_{j}^{K}
\label{SSS}
\end{eqnarray}
$\langle I_j \rangle_0$ and $\tilde{S}_{j}^0$ invoke only the
tunneling operators $T_j$ and $T_j^{\dagger}$ between edges $0$ and
$j$.  As discussed above, these terms do not involve 
Klein factors as they couple only two edges, and 
they thus correspond exactly to the current and noise in a two 
terminal system derived in 
\cite{fendley_ludwig_saleur}:
\begin{eqnarray}
\langle I_{j} \rangle_0&\equiv& -ev_F
\int_{-\infty}^{+\infty} d\theta \rho(\theta)
\vert S_{+-}(\theta-\theta_B)\vert^2
\nonumber
\\
\tilde{S}_{j}^0&\equiv&
\frac{\nu e}{2(1-\nu)}
\left[ V \frac{d\langle I_{j} 
\rangle_0}{dV}-\langle I_{j} \rangle_0\right]
\label{exact_solution}
\end{eqnarray}
with $\rho(\theta)$ the kink density, 
$S_{+-}(\theta-\theta_B)$ the kink transmission  
probability and the barrier height appears
via the kink parameter $\theta_B$.     
Its leading order 
is proportional to $\vert \Gamma_j \vert^2$, the Poisson limit.   
On the other hand, the second contribution of 
Eq.(\ref{SSS})
involves additional tunneling processes 
between the injecting edge  $0$ and other edges $k$, for instance $ k> j$. 
In perturbation theory, 
these additional terms are proportional to 
$\vert \Gamma_j^{p} \Gamma_k^{p'} \vert ^2$ ($p,p'\geq 1$) and they 
signal the possibility that quasiparticles 
from an external ``reservoir'' (edge $k$)
are exchanged with the injecting edge $0$.
The leading order $\vert \Gamma_j\Gamma_k\vert^2$ correction to
the Klein factor  contribution reads:  
\begin{eqnarray}
&&\tilde{S}_{j}^{K} (\omega = 0) = \frac{2 \vert \Gamma_j  \Gamma_k \vert ^2}{h^4 \alpha^4}\; 
\sum_{\eta \eta_1  \eta_2} \sum_{\epsilon\epsilon'}  
\eta_1 \eta_2 \int\! dt\!\int\! dt_1 \!\int\! 
dt_2  \nonumber \\ 
 &&  e^{i\omega_0 \left[\epsilon t  +\epsilon' (t_1 - t_2)\right]} 
e^{2 \nu \left[G_{\eta
-\eta}(0,t)+G_{\eta_1 \eta_2}(0,t_1-t_2)
\right]}  \nonumber
\\
&& \times 
\Big( \frac{e^{ \nu\epsilon\epsilon'\left[G_{\eta\eta_2}(-a,t-t_2) 
+G_{-\eta\eta_1}(-a,-t_1)\right]}}{e^{\nu\epsilon\epsilon' 
\left[G_{\eta \eta_1}(-a,t-t_1) 
+G_{-\eta\eta_2}(-a,-t_2) 
\right]}}  e^{i \xi'} - 1  \Big)
\end{eqnarray}

Where the Klein factor time ordered product reads:
\begin{eqnarray} 
e^{i \xi'} &=& \langle T_K 
e^{-i\epsilon\sqrt{\nu}\theta_j(t^{\eta})} 
e^{i\epsilon\sqrt{\nu}\theta_j(0^{-\eta})} 
e^{-i\epsilon'\sqrt{\nu}\theta_k(t_1^{\eta_1})} 
e^{i\epsilon'\sqrt{\nu}\theta_k(t_2^{\eta_2})} \rangle 
\nonumber\\
&=&
e^{\nu \pi}
e^{ i \nu \pi \epsilon\epsilon' 
\left(  
(\eta +\eta_2)sgn(t-t_2)
+(-\eta +\eta_1)sgn(-t_1)\right)/4}
\nonumber\\
&&
e^{ i \nu \pi \epsilon\epsilon' 
\left(  -(\eta +\eta_1)sgn(t-t_1)
+(\eta -\eta_2)sgn(-t_2)
\right)/4}
\end{eqnarray}
The calculation of the current and its contribution 
$\langle I_1 \rangle_K$
linked to the Klein factors follows the same guidelines,
except that the Keldysh evolution operator in Eq. (\ref{current_keldysh})
is expanded to third order. 

In conclusion, a systematic procedure is provided for
computing  transport quantities such as current, noise, noise
correlations in multiple FQHE edges. Generalizations 
to hierarchical fractional Hall states and to 
multi-quasiparticle tunneling are envisioned.   
Multiple quantum Hall edge transport problems are 
necessary to address the nature of the statistics of the 
underlying excitations: 
the computation of the noise cross correlations requires Klein
factors.   Multiple edges allow to study fractional quantum
transport through a constriction  of a prepared fluctuating
injecting current  \cite{private_heiblum}. This injecting
current  fluctuates because of passage through a 
prior gate, and the whole system requires once again three edges.
 
The current and the noise contain an
``exact'' contribution where all other tunneling processes
are  discarded, together with Klein factor corrections.
This latter contribution illustrates how current and 
noise are affected by the presence of additional stochastic
contributions.
It is symptomatic of a tunneling process ($T_2$) which
is influenced by another via a non linear coupling ($T_1$): in
Fig. \ref{conect}, the fluctuations at $1$ propagate to the right
along the bottom edge, affecting  all transport properties
at $2$. In the language of quantum dissipative systems, it can be
mapped to the problem of a brownian motion \cite{schmid} in a
periodic potential with two coupled particles.   

T.M. acknowledges support from the LSF program at 
the Brown Submicron Center of Weizmann Institute.

\end{multicols}
\end{document}